\documentclass[nonacm,sigconf]{acmart}
\usepackage{pifont}
\usepackage{algorithm}
\usepackage{algpseudocode}
\usepackage{listings}
\usepackage{xcolor}
\usepackage{multirow}
\usepackage{enumitem}
\usepackage{beramono} 
\usepackage{hyperref}
\usepackage{url}
\usepackage[skip=1pt]{caption}
\usepackage[aboveskip=1pt]{subcaption}

\definecolor{codegreen}{rgb}{0,0.6,0}
\definecolor{codegray}{rgb}{0.5,0.5,0.5}
\definecolor{codepurple}{rgb}{0.58,0,0.82}
\definecolor{backcolour}{rgb}{0.95,0.95,0.92}
\definecolor{deepblue}{rgb}{0,0,0.5}
\definecolor{darkgray}{rgb}{0.4,0.4,0.4}
\definecolor{listingbackground}{rgb}{0.98,0.98,0.98}
\definecolor{highlightmin}{RGB}{141,210,197}
\definecolor{highlightsecondmin}{RGB}{186,227,220}
\definecolor{highlightsecondmax}{RGB}{246,179,172}
\definecolor{highlightmax}{RGB}{255, 180, 170}
\definecolor{rulecolor}{rgb}{0.8,0.8,0.8} 

\lstdefinestyle{sqlstyle}{
    language=SQL,
    backgroundcolor=\color{listingbackground}, 
    commentstyle=\color{codegreen}\itshape,    
    keywordstyle=\color{deepblue}\bfseries,   
    stringstyle=\color{codepurple},
    numberstyle=\tiny\color{darkgray},
    basicstyle=\ttfamily\footnotesize,      
    breakatwhitespace=false,
    breaklines=true,
    captionpos=b,                           
    keepspaces=true,
    numbers=left,
    numbersep=5pt,                          
    showspaces=false,
    showstringspaces=false,
    showtabs=false,
    tabsize=2,
    frame=tb,                               
    framerule=0.5pt,                        
    rulecolor=\color{rulecolor},
    title=\lstname,                         
    escapeinside={\%*}{*)},                 
    morekeywords={AS, REVENUE, DATE, LIMIT} 
}

\AtBeginDocument{%
  }

\setcopyright{acmlicensed}
\copyrightyear{2018}
\acmYear{2018}
\acmDOI{XXXXXXX.XXXXXXX}
\acmConference[Conference acronym 'XX]{Make sure to enter the correct
  conference title from your rights confirmation email}{June 03--05,
  2018}{Woodstock, NY}
\acmISBN{978-1-4503-XXXX-X/2018/06}

\begin{document}

\title{CARPO: Leveraging Listwise Learning-to-Rank for \underline{C}ontext-\underline{A}wa\underline{r}e Query \underline{P}lan \underline{O}ptimization}

\author{Wenrui Zhou}
\affiliation{%
  \institution{Beijing Institute of Technology}
  \city{Beijing}
  \country{China}
  }
\email{chouwilliam422@gmail.com}

\author{Qiyu Liu\authornotemark[1]}
\affiliation{%
  \institution{Southwest University}
  \city{Chongqing}
  \country{China}
}
\email{qyliu.cs@gmail.com}
\authornote{Corresponding author.}

\author{Jingshu Peng}
\affiliation{%
  \institution{HKUST}
  \city{Hong Kong SAR}
  \country{China}
}
\email{jpengab@cse.ust.hk}

\author{Aoqian Zhang}
\affiliation{%
  \institution{Beijing Institute of Technology}
  \city{Beijing}
  \country{China}
}
\email{aoqian.zhang@bit.edu.cn}

\author{Lei Chen}
\affiliation{%
  \institution{HKUST}
  \city{Hong Kong SAR}
  \country{China}
}
\email{leichen@cse.ust.hk}

\renewcommand{\shortauthors}{Paper ID: 1039}

\begin{abstract}
Efficient data processing is increasingly vital, with query optimizers playing a fundamental role in translating SQL queries into optimal execution plans. 
Traditional cost-based optimizers, however, often generate suboptimal plans due to flawed heuristics and inaccurate cost models, leading to the emergence of Learned Query Optimizers (LQOs). 
To address challenges in existing LQOs, such as the inconsistency and suboptimality inherent in pairwise ranking methods, we introduce CARPO, a generic framework leveraging listwise learning-to-rank for context-aware query plan optimization. 
CARPO distinctively employs a Transformer-based model for holistic evaluation of candidate plan sets and integrates a robust hybrid decision mechanism, featuring Out-Of-Distribution (OOD) detection with a top-$k$ fallback strategy to ensure reliability. 
Furthermore, CARPO can be seamlessly integrated with existing plan embedding techniques, demonstrating strong adaptability. 
Comprehensive experiments on TPC-H and STATS benchmarks demonstrate that CARPO significantly outperforms both native PostgreSQL and Lero, achieving a Top-1 Rate of \textbf{74.54\%} on the TPC-H benchmark compared to Lero's 3.63\%, and reducing the total execution time to 3719.16 ms compared to PostgreSQL's 22577.87 ms.
\end{abstract}





\maketitle

\section{Introduction}
As modern data continues to grow in both volume and complexity, efficient data management and processing have become increasingly crucial. 
A core component in addressing this challenge is the query optimizer\cite{chaudhuri1998overview}, which translates SQL queries into efficient physical execution plans, aiming to minimize processing time. 

To manage the exponentially large candidate execution plan space, practical DBMSs (e.g., PostgreSQL\cite{pg} and MySQL\cite{mysql}) usually adopt cost-based optimizers (CBO). 
Though developed and used for decades, CBOs are observed to frequently generate suboptimal execution plans as, 
\ding{182} rule-based heuristics are likely to remove optimal plans from the candidate pool\cite{mekchay2017limitations}, and~\ding{183} cost models frequently produce inaccurate estimates due to the use of predefined magic constants and sometimes questionable statistical assumptions for scoring and ranking candidate plans\cite{armbrust2020wrong}. 


\begin{figure*}
\centering
\includegraphics[width=0.75\linewidth]{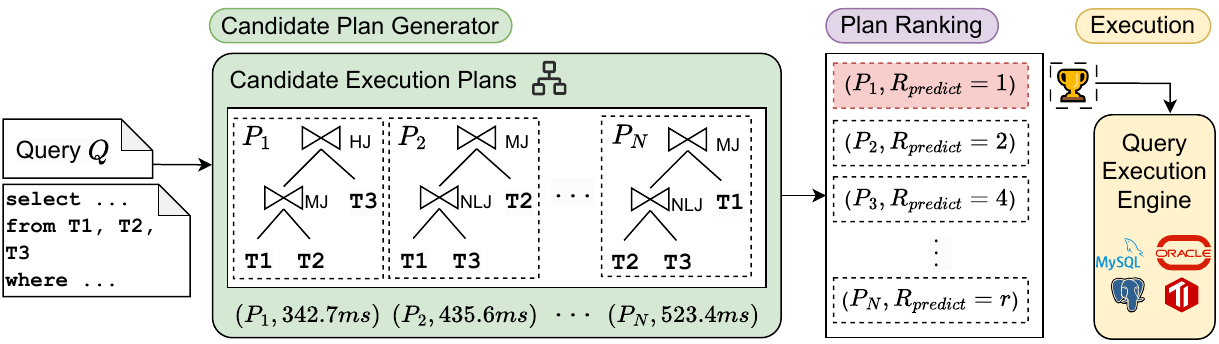}
\caption{
Illustration of  the generic framework of a learned query optimizer (LQO).  
Given a query, the LQO generates multiple candidate execution plans, ranks them based on predicted performance, and then executes the plan deemed optimal by the learned model within the query execution engine (e.g., MySQL, Oracle).
}
\label{fig:general_mqo}
\vspace{-2ex}
\end{figure*}

In response to the inherent limitations of CBOs, learned query optimizers (LQOs), a new paradigm driven by advancements in machine learning (ML), has emerged\cite{krishnan2018learning, trummer2019skinnerdb, marcus2021balsa}.
Seeking to overcome the reliance on potentially inaccurate, handcrafted cost models and heuristics, LQOs leverage ML models trained on query execution data or database characteristics to either predict plan performance or directly determine the most efficient execution strategy\cite{marcus2019neo, marcus2019bao}.


As illustrated in Figure~\ref{fig:general_mqo}, the task of \textbf{learning to optimize queries} centers on leveraging data-driven models to improve the selection of query execution plans.
It involves training models that generalize from observed plan behaviors and execution feedback to make better optimization decisions for unseen queries\cite{sun2020end}. 
Rather than hard-coding the rules for plan selection, learned approaches treat query optimization as a learning problem, where the model jointly considers the plan-level features (e.g., join algorithms) and data-level features (e.g., number of distinct values in a column). 
This shift enables the optimizer to adapt diverse data distributions, workload patterns, and system environments\cite{hilprecht2019deepdb}, which are capabilities that conventional optimizers often lack. 

Existing LQOs generally adopt one of two main strategies:
\ding{182} \textbf{Predict absolute plan execution metrics}, such as the actual cost or latency. 
Optimizers in this category, including notable examples like Neo\cite{marcus2019neo}, Balsa\cite{marcus2021balsa} and Bao\cite{marcus2019bao}, train models to estimate these absolute values. 
For instance, Neo uses a Tree Convolution Neural Network (Tree-CNN) to predict optimal expected execution latency, while Bao also uses a Tree-CNN to estimate plan latency under different hint sets, employing Thompson sampling for adaptive learning. 
While comprehensive, accurately modeling all factors influencing query execution is complex, potentially impacting model stability and efficiency. 
Recognizing this, \ding{183} focuses on \textbf{predicting the relative performance or rank} among different plans. 
Instead of exact values, these methods often learn pairwise comparisons. 
Lero\cite{zhu2023lero}, for example, trains a binary classifier to determine which of two plans ($P_1$ and $P_2$) will perform better, simplifying the learning task compared to absolute cost prediction.

Despite their advantages in simplifying the learning task and improving prediction robustness, pairwise approaches are not without limitations.
Inspired by research in Information Retrieval (IR), two critical drawbacks are identified in this approach. 
\ding{182}~\textbf{Inconsistency:} 
A focus on optimizing local comparisons between pairs of plans does not inherently guarantee a globally consistent ranking order. 
This lack of enforced transitivity means that scenarios involving cyclical preferences (e.g., $P_1\succ P_2\succ P_3\succ P_1$) can arise, making the selection of a best plan less definitive.
\ding{183}~\textbf{Sub-optimality:}
The typical pairwise training objective centers on minimizing the relative error in ordering pairs, rather than directly optimizing for the absolute performance quality of the highest-ranked plan. As a result, the selected plan might merely be the dominant one in pairwise matchups, rather than the genuinely optimal execution strategy available for the query.

Furthermore, a crucial characteristic observed in query optimization is that the performance difference among the top few candidate plans (e.g., the top-3 or top-5) for a given query can often be negligible. 
This implies that rigidly focusing solely on identifying the single, absolute best plan might be unnecessarily restrictive and potentially sensitive to minor prediction inaccuracies.
An effective optimizer should ideally identify a set of high-quality candidate plans, offering both flexibility and robustness.

To overcome the potential inconsistency and sub-optimality inherent in pairwise methods, and to effectively handle scenarios with multiple near-optimal plans, we propose CARPO, a listwise ranking approach. 
The \textit{Candidate Plan Generator} first enumerates alternative execution plans for a given query, which are encoded into vector representations by the \textit{Plan Embedder}. 
At the core, the \textit{Listwise Ranking Predictor} employs a Transformer to jointly evaluate all plan embeddings, capturing inter-plan dependencies to produce a holistic ranking. 
Finally, the \textit{Hybrid Decision Maker} selects the top-ranked plan or defers to the native CBO plan when model confidence is low, ensuring both effectiveness and reliability.

Compared to existing LQOs, CARPO offers a more holistic alternative to address drawbacks mentioned above for the following three reasons:

\noindent
\ding{182} \textbf{Global Context Integration: }
By evaluating the entire list of candidate plans simultaneously, listwise models can capture the global context and inter-plan dependencies\cite{vaswani2017attention}. 
This comprehensive awareness ensures a more consistent and transitive overall ranking, directly addressing the inconsistency issues of pairwise comparisons.

\noindent
\ding{183} \textbf{Direct Optimization for Top-Rank Quality: }
Unlike pairwise methods that focus on relative errors, listwise loss functions are defined over the full ranking\cite{cao2007learning}. 
This enables the model to be trained specifically to optimize the absolute quality and correct positioning of the best plans within the list aligning well with the goal of identifying a high-quality set even if the absolute top-1 is ambiguous or has close competitors.
As a result, it better aligns with the ultimate goal of identifying the genuinely optimal execution strategy rather than just the ``least flawed'' one.

\noindent
\ding{184} \textbf{Leveraging Rich Supervision:} 
While pairwise methods are effective in IR domains like sparse recommendation where labeled data is limited, query optimization workflows often involve the generation and performance evaluation of numerous candidate plans. 
This provides richer supervisory signals, which listwise approaches are inherently designed to utilize more effectively for model training compared to methods reliant on sparse pairwise labels.

We summarize our technical contributions as follows.
\begin{enumerate}[leftmargin=*]
\item
\textbf{We introduce CARPO}, a Transformer-based listwise learning-to-rank framework that captures global context across candidate query plans and incorporates a robust top-$k$ fallback strategy with OOD detection for reliability.

\item
\textbf{CARPO outperforms both PostgreSQL and SOTA LQOs:} On TPC-H, it reduces execution time by up to 83.5\% over PostgreSQL and achieves 74.5\% top-1 accuracy, versus 3.6\% for Lero.

\item
\textbf{CARPO is adaptable and robust:} It is a plug-and-play framework by supporting diverse plan embedders (e.g., TreeLSTM), and its top-$k$ strategy effectively mitigates model uncertainty, as confirmed by ablations and case studies.
\end{enumerate}

\section{Related Work}

\subsection{Learning to Rank (LTR)}
Learning to Rank (LTR) encompasses supervised machine learning techniques designed to construct ranking models, primarily originating from Information Retrieval\cite{liu2011learning}. 
LTR algorithms are typically categorized into three main approaches based on their input representation and loss function:

\noindent\underline{\textbf{Pointwise Approach.}} 
This approach treats each item (e.g., document, query plan) independently, framing the task as a standard regression or classification problem to predict an absolute relevance score or class for each item. 
While simple, it fundamentally ignores the relative order crucial for ranking and optimizes loss functions often misaligned with standard list-based evaluation metrics like NDCG\cite{liu2011learning}.

\noindent\underline{\textbf{Pairwise Approach.}}
This approach focuses on the relative order between pairs of items, reformulating ranking as a binary classification problem on item pairs. 
This is conceptually closer to ranking than the pointwise approach. 
Prominent algorithms include RankSVM\cite{herbrich2000large} , RankBoost\cite{freund2003efficient} , and RankNet\cite{burges2005learning}. 
However, standard pairwise methods optimize pairwise accuracy, which is an indirect alternative for listwise metrics, and can suffer from high computational complexity due to the quadratic number of pairs. 
Extensions like LambdaRank\cite{burges2006learning}  and LambdaMART\cite{wu2010adapting} attempt to bridge this gap by incorporating gradients from listwise metrics into the pairwise framework.

\noindent\underline{\textbf{Listwise Approach.}} 
This approach directly addresses the limitations of the previous two by treating the entire list of items associated with a query as a single instance for learning. 
Listwise methods aim to optimize loss functions that are directly derived from or approximate standard list-based evaluation metrics. Examples include ListNet\cite{cao2007learning}  and ListMLE\cite{xia2008listwise}. 
By considering the positions and interdependencies of items within the entire list, listwise methods offer a more direct way to optimize ranking quality. 
This paper adopts a listwise LTR approach to learn context-aware query plan preferences, aiming to directly optimize the quality of the chosen plan within the set of alternatives.

\subsection{Learning to Optimize Queries}
Machine learning (ML) techniques have been increasingly applied to various facets of query optimization. These efforts can be broadly categorized based on whether they predict absolute metrics of query plan execution or learn the relative quality or rank of plans.

\noindent\underline{\textbf{Learning Absolute Plan Execution Metrics.}} 
This category contains methods that learn to predict an absolute score or specific metric for individual query plans or their constituent parts, which are then often used by an optimizer.
A significant portion of research focuses on \textbf{learned cardinality estimation}, where models predict cardinalities for subqueries based on their features\cite{kipf2019learned} or learn data distributions to derive these estimates\cite{hilprecht2019deepdb, yang2019deep}. 
Similarly, learned cost models aim to \textbf{predict the execution cost} of a given query plan more accurately than traditional heuristics\cite{sun2020end}. 
Some end-to-end learned query optimizers, like Neo\cite{marcus2019neo}, also utilize learned models for cost prediction or value function approximation to assign scores to candidate plans, while systems like Bao\cite{marcus2019bao} learn to score the utility of different hint configurations. 
These pointwise methods treat each evaluation as an independent prediction, akin to regression or classification as seen in pointwise Learning to Rank\cite{liu2011learning}, but often don't directly optimize for the final ranking quality among a set of candidate plans.

\noindent\underline{\textbf{Learning Relative Plan Quality and Rank.}} 
This group of techniques shifts from predicting absolute scores to learning the relative desirability or rank between pairs of query plans or optimization choices, aiming to learn a preference function. 
Many works model join order selection or other plan construction tasks using reinforcement learning, where an agent learns a policy by implicitly comparing the expected outcomes of different actions, thereby learning relative preferences between resulting states or partial plans\cite{krishnan2018learning, marcus2018deep, ortiz2018learning, yu2020learning}. 
Systems like SkinnerDB\cite{trummer2019skinnerdb} also learn relative preferences by choosing between alternative execution strategies. 
While these pairwise approaches capture relative quality, their optimization objective is often an approximation of global ranking quality and they typically do not consider the full context of the entire list of candidate plans, contrasting with listwise approaches in our work.

\section{System Overview}

CARPO is a learned query optimization framework designed for reliable and context-aware plan selection, leveraging a listwise learning-to-rank approach. 
Unlike pointwise methods that score plans independently or pairwise methods that compare plans in pairs, CARPO evaluates the relative quality of candidate plans within the full context of the query's potential execution space. 
This holistic evaluation aims to overcome the ranking inconsistencies and suboptimality challenges inherent in methods that rely solely on local comparisons. 
By considering the entire list of plans together, CARPO captures global dependencies and structural relationships crucial for accurate ranking.

\begin{figure*}[ht]
\centering
\includegraphics[width=0.9\linewidth]{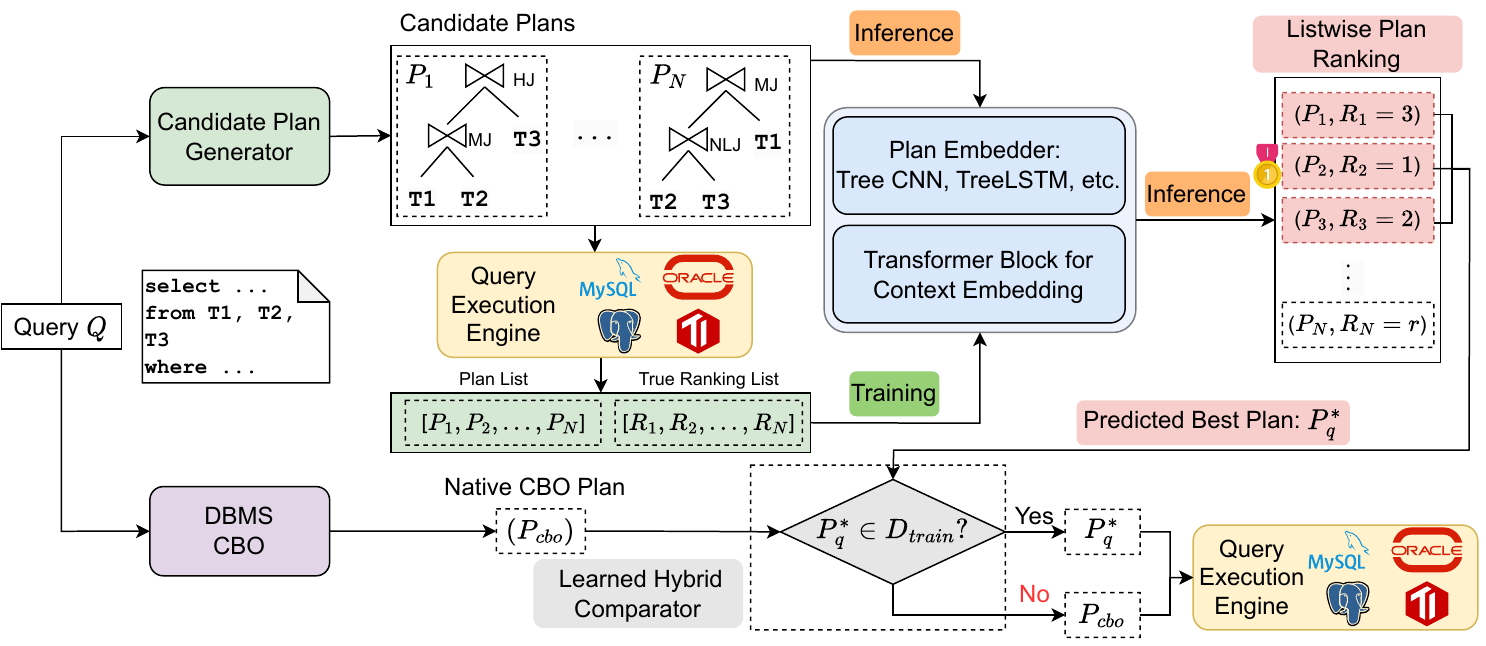}
\caption{
Illustration of CARPO's system overview. 
During training, CARPO learns to rank candidate plans generated for a query using a Transformer-based model, leveraging execution costs from the Query Execution Engine. 
At inference, CARPO selects the learned top plans or falls back to the DBMS's Cost-Based Optimizer (CBO) if the learned model's prediction is deemed uncertain, ensuring robust performance.
}
\label{fig:overview}
\end{figure*}

Figure~\ref{fig:overview} illustrates the overall architecture of CARPO. 
For an input query Q, a set of candidate execution plans \( \{P_1, P_2, \cdots, P_n \}\) is generated. 
These plans then flow through the core components of CARPO. 
First, the \textit{\textbf{Data Collector}} module gathers plan details and performance metrics, often involving physical execution to establish ground truth labels. 
Next, the \textit{\textbf{Plan Embedder}} transforms each plan's structural and cost information into a rich feature representation suitable for machine learning models. As a general framework, researchers have the flexibility to select plan embedding modules, such as Tree Convolutional Neural Networks (Tree-CNNs)\cite{marcus2019neo} or Tree Long Short-Term Memory (Tree-LSTM) networks\cite{tai2015improved}.
The core \textit{\textbf{Listwise Ranking Predictor}} then takes the embeddings for the entire set of plans \( \{P_1, P_2, \cdots, P_n \}\) and predicts a ranked order, often employing sequence models like Transformers to capture inter-plan context. 
Finally, to ensure reliability, especially for novel or out-of-distribution queries, the \textit{\textbf{Hybrid Decision Maker}} assesses the model's prediction confidence and intelligently decides whether to utilize the plan recommended by CARPO or to fallback to the database's native cost-based optimizer. 
These components work in concert to select high-quality execution plans based on learned, context-aware rankings.

\noindent\underline{\textbf{Data Collector.}} 
The foundation of CARPO relies on high-quality training data that captures real-world plan performance.
Our data collection process involves three main steps: \ding{182} \textbf{Plan Exploration.} 
For a given query, the \textbf{plan exploration} module generates a diverse set of candidate execution plans \( \{P_1, P_2, \cdots, P_n \}\).
We adopt a strategy similar to Lero, perturbing internal statistics (e.g. cardinality estimates) before feeding them into the native query optimizer. 
This technique encourages the optimizer to explore alternatives beyond its default choice, producing a varied set of plans that includes potentially superior options as well as sub-optimal ones. 
The goal is to generate candidates that are not only potentially high-performing but also sufficiently diverse to train a robust ranking model effectively.
\ding{183} \textbf{Physical Execution.} 
During the offline training phase, these generated candidate plans undergo \textbf{physical execution} on Postgres database system. 
We execute each plan multiple times to mitigate transient factors like cache state or system load, and record their actual execution latencies. 
These measured latencies serve as the crucial ground truth signal, reflecting the true performance differences between plans.
\ding{184} \textbf{Ranking Organization.} 
The collected data is structured for listwise learning through \textbf{ranking organization}. 
For each query \(Q\), its corresponding candidate plans \(\{P_1 ,\cdots,P_n \}\) are sorted based on their average measured execution times. 
This process creates a ground truth ranked list \(r_Q =\{r_1 ,r_2 ,\cdots,r_n\}\), where \(r_i\) indicates the true performance rank of plan \(P_i\) among the candidates for that query. 
These plan sets and their associated ground truth rankings form the training dataset \((P_Q ,r_Q)\) used to train the listwise ranking predictor.

\noindent\underline{\textbf{Plan Embedder.}} 
Once candidate plans are available, the \textit{Plan Embedder} component translates their structural and feature information into numerical vector representations suitable for the subsequent ranking model. 
This typically involves processing the plan tree and its associated features (e.g., operator types, estimated costs, cardinalities) using techniques such as tree-based neural networks (e.g. TreeCNN, TreeLSTM, etc.) or other feature extraction methods.

Crucially, CARPO is designed with ideas of modularity and plug-and-play. 
While effective embeddings are necessary for the predictor to function, the specific embedding technique is not the core focus of our work. 
This component is intentionally flexible, allowing researchers to substitute different plan embedding strategies as needed or as new methods emerge.

\noindent\underline{\textbf{Listwise Ranking Predictor.}} 
At the heart of CARPO lies the Listwise \textit{Ranking Predictor}. 
This component takes the sequence of plan embeddings \(\{h_{p1} ,h_{p2} ,\cdots,h_{pn} \}\) generated by the \textit{Plan Embedde}r for a given query \(Q\). 
Unlike traditional pairwise approaches that learn to compare two plans in isolation, our predictor adopts a listwise strategy, evaluating the entire set of candidate plans simultaneously.

This holistic approach allows the model to capture the crucial query-level context and inter-plan dependencies often missed by methods focusing only on local comparisons. 
By considering all plans together, the model can better understand the relative trade-offs and produce more globally consistent rankings. 
We employ Transformer encoder architecture which utilizes self-attention mechanisms to effectively model the relationships and dependencies across the full list of plan embeddings. 
The model is trained end-to-end using position-aware cross-entropy, a typical listwise loss function.
It directly optimizes the accuracy of the predicted rank positions for all plans within the list relative to their ground truth performance ranks obtained during data collection phase. 
This enables CARPO to learn a nuanced ranking function that leverages the collective information present in the set of candidate plans.

\noindent\underline{\textbf{Hybrid Decision Maker.}} 
While the listwise \textit{Ranking Predictor} aims to identify superior execution plans, learned models can sometimes produce unreliable predictions, especially for queries that differ significantly from the training data (Out-Of-Distribution, or OOD queries).
To ensure consistent performance and robustness, CARPO incorporates a \textit{Hybrid Decision Maker} module. 
The primary purpose of this component is to act as a safety net, leveraging the strengths of both the learned model and the database's native CBOs.

The core principle involves assessing the confidence of the ranking produced by the listwise \textit{Ranking Predictor}. 
This is often achieved through an OOD detection mechanism that identifies potentially problematic queries where the learned model might struggle. 
If the predictor's confidence for its top-ranked plans fall below a certain threshold, or if the query is flagged as OOD, the \textit{Hybrid Decision Maker} bypasses the learned plan. 
Instead, it defaults to using the execution plan generated by the system's traditional CBO. 
This fallback strategy ensures that CARPO maintains a reliable baseline performance provided by conventional CBO, while still capitalizing on the potentially superior plans identified by the learned model for queries it can confidently handle.
The detailed mechanisms will be explained in Section ~\ref{sec:hybrid-decision-making}.

\section{A Learned Context-Aware Query Plan Optimizer}
Following the system overview, this section elaborates on the core predictive component of CARPO: the learned model designed to produce context-aware, listwise rankings of candidate query execution plans. 
As outlined previously, achieving accurate plan selection requires not only understanding individual plan characteristics but also incorporating the relative merits of plans within the full set generated for a query. 
To this end, CARPO utilizes a neural architecture specifically tailored for listwise learning-to-rank. 
We first detail this model's architecture, covering: \ding{182} \textbf{Plan Embedding Module}, which converts the structural and statistical features of each plan into a vector representation; \ding{183} \textbf{Contextual Ranking with Transformers}, which processes the sequence of plan embeddings to capture inter-plan dependencies and ranks accordingly; \ding{184} \textbf{Listwise Loss Function} employed to directly optimize the ranking quality. 
On top of that, we describe the hybrid approach used for test-time plan selection.

\subsection{Model Design}
\begin{figure*}[ht]
\centering
\includegraphics[width=0.78\linewidth]{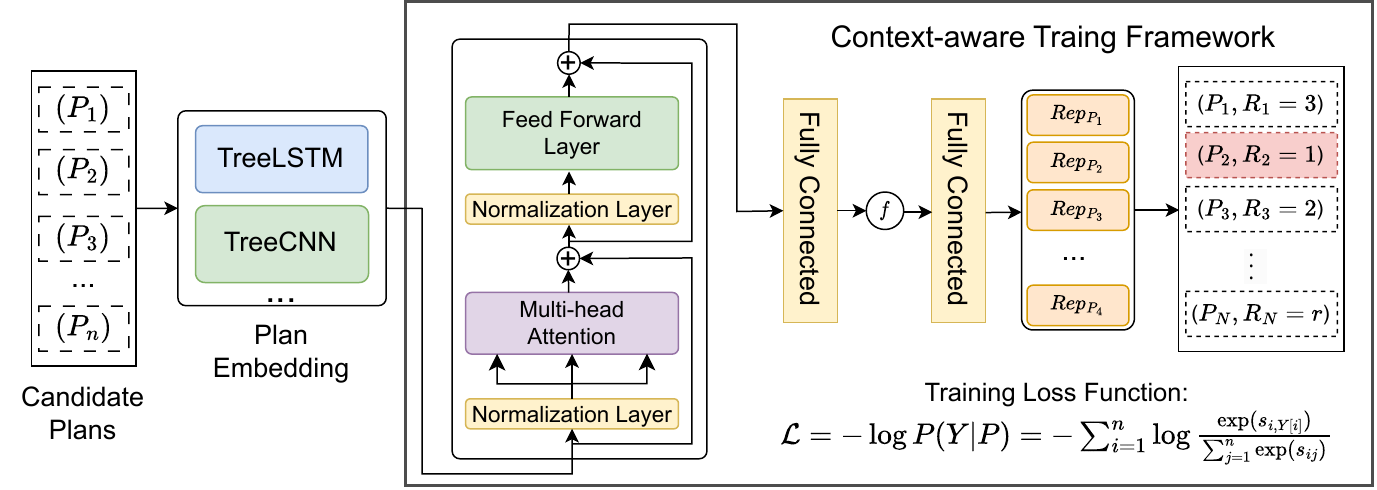}
\caption{
Illustration of CARPO's model architecture within the context-aware training framework. 
Candidate plans are embedded using a flexible Plan Embedder (TreeLSTM or TreeCNN in our case), and these embeddings are processed by a Transformer encoder with multi-head attention and feed-forward layers. 
The model is trained to predict plan rankings, minimizing a listwise loss function that directly optimizes the probability of the ground truth ranking.
}
\label{fig:model}
\end{figure*}

The neural network architecture employed by CARPO is specifically designed for the listwise ranking task. 
It comprises integrated components responsible for encoding individual plan structures and subsequently capturing the essential context across the set of candidate plans for a given query. 
Figure~\ref{fig:model} provides an overview of this architecture, the details of which are described in the following parts.

\noindent\underline{\textbf{Plan Embedding Module.}}
The first crucial step in CARPO's model architecture is the Plan Embedding Module. 
Its primary goal is to transform each complex, tree-structured query execution plan \(P_i\) from the candidate set \(\{P_1, \cdots, P_n\}\) into a dense, fixed-size vector representation, or embedding \(h_{P_i}\). 
This numerical representation serves as the input for the subsequent context-aware ranking layers.

Methodologically, CARPO is designed for flexibility, allowing different neural network architectures capable of processing tree-structured data to serve as the embedding model \(\text{PlanEmb}\). 
While techniques like TreeCNNs or TreeLSTM networks are common choices, the framework is not tied to a single specific embedder. 
This modularity allows CARPO to adapt easily to different types of plan features or leverage future advancements in representing structured data.

The core design principle, regardless of the specific embedder chosen, is to effectively capture the critical information inherent in the query plan. 
This includes encoding details about the physical operators used (e.g., Hash Join, Index Scan), associated estimated properties like cardinalities and costs, and involved tables and predicates. 
It should also capture the hierarchical structure defining the flow of data through the plan.
Technically, this involves featurizing each node (operator) in the plan tree with relevant attributes (e.g., operator type as one-hot encoding, normalized log-cardinality estimates, table identifiers). 
The chosen tree-based network (like TreeCNN) then processes this featurized tree, often using recursive or convolutional operations, to aggregate information and produce a final vector embedding for the entire plan.

It's important to distinguish CARPO's embedding goal from models focused solely on predicting absolute plan latency or cost. 
While the learned embedding \(h_{P_i}\) should implicitly encode factors relevant to plan performance, its primary purpose within CARPO is to serve as a rich input for the downstream listwise ranking model. 
The focus is on creating representations that enable effective comparison and ranking within the context of other plans, rather than precisely matching a specific execution time value.

This approach offers advantages in flexibility and potentially simplifies training. 
By decoupling the embedding from the need to predict exact costs, the model can focus on learning representations optimized for relative ranking. 
Furthermore, the modular design ensures that CARPO can readily incorporate improved embedding techniques as they become available, which enhances its representational power without altering the core listwise ranking mechanism.

\noindent\underline{\textbf{Contextual Ranking with Transformers.}}
Central to CARPO is its ability to perform context-aware ranking by considering the entire list of candidate plans \(\{P_1,P_2,\cdots,P_n\}\) simultaneously. 
This is achieved using a Transformer-based architecture that processes the sequence of plan embeddings \(\{h_{p_1},h_{p_2},\cdots,h_{p_n}\}\) generated by the Plan Embedding Module described previously.
It mainly consists two stages: 
\ding{182} \textbf{Integrating Context via Self-Attention.}
The key mechanism for integrating context is the multi-head self-attention layer within the Transformer encoder\cite{vaswani2017attention}.
For a given plan \(P_i\) represented by embedding \(h_{p_i}\), self-attention computes how relevant every other plan \(P_j\) (represented by \(h_{p_j}\)) in the candidate list is to \(P_i\).
Mathematically, this relevance is often captured using scaled dot-product attention scores \(A_{ij}\) calculated as:
\begin{equation}
    A_{ij} = \frac{\exp( (h_{p_i} W^Q) (h_{p_j} W^K)^T / \sqrt{d_k} )}{\sum_{k=1}^{n} \exp( (h_{p_i} W^Q) (h_{p_k} W^K)^T / \sqrt{d_k} )},
\end{equation}
where \(W^Q\), \(W^K\) are learned weight matrices projecting embeddings into Query and Key spaces respectively, and \(d_k\) is the dimension of the Key vectors.
The contextualized representation \(z_i\) for plan \(P_i\) is then computed as a weighted sum of value representations (derived from \(h_{p_j}\) using another weight matrix \(W^V\)) of all plans in the list, weighted by these attention scores:
\begin{equation}
    z_i = \sum_{j=1}^{n} A_{ij} (h_{p_j} W^V).
\end{equation}
This mechanism ensures that the final representation \(z_i\) for each plan is not determined in isolation but is explicitly contextualized by its relationship with all other candidate plans for the query.

\ding{183} \textbf{Listwise Ranking Formulation.}
CARPO employs a listwise learning-to-rank (LTR) approach, processing the entire set of candidate plans \(P_q = \{p_1, ..., p_n\}\) simultaneously to predict their optimal ranking. 
Given the contextualized plan embeddings \(Z = \{z_1, ..., z_n\}\) from the Transformer encoder, the model effectively treats ranking as a sequence-level classification problem. 
Instead of predicting isolated scores or pairwise preferences, it learns to produce scores \( s_{ij} \) that reflect the suitability of assigning plan \( p_i \) to rank position \( j \).
This allows the model to directly consider the relative positioning of all plans within the list.

The core objective is to learn the parameters \(\Theta\) of a function \(f_{\Theta}\) that generates these scores \(s_{ij}\) such that the resulting predicted ranking \(\pi_{\Theta}\) maximizes a ranking utility function \(U(\pi)\) that evaluates the quality of a given ranking. 
Formally, 
\begin{equation}
    \pi_{\Theta} = \arg \max_{\pi \in \Pi_n} U(\pi | f_{\Theta}(P_q)),
\end{equation}
where \(\Pi_n\) is the set of all permutations. 
The specific loss function used during training (detailed next) aims to align the predicted scores \(s_{ij}\) with the ground truth ranking, thereby optimizing the quality of the entire ranked list directly, rather than focusing only on individual plan scores or local comparisons.

\noindent\underline{\textbf{Listwise Loss Function.}}
The parameters $\Theta$ of the CARPO model are trained by minimizing a listwise loss function designed to directly optimize the ranking quality. 
Specifically, we adopt a position-aware cross-entropy loss, which measures the difference between the predicted ranking scores and the ground truth ranking for each query's candidate plan list.

Recall that the model produces scores \(s_{ij}\) representing the predicted relevance of assigning plan \(p_i\) to rank position \(j\). 
Let \(y_i\) denote the ground-truth rank position of plan \(p_i\) within the list of \(n\) candidates (derived from actual performance metrics). The loss function \(\mathcal{L}\) for a single query's list is defined as:
\begin{equation}
    \mathcal{L} = - \sum_{i=1}^{n} \log P(y_i | p_i, P_q) = - \sum_{i=1}^{n} \log \frac{\exp(s_{i, y_i})}{\sum_{j=1}^{n} \exp(s_{ij})}.
\end{equation}
Here, the term \(\frac{\exp(s_{i, y_i})}{\sum_{j=1}^{n} \exp(s_{ij})}\) calculates the model's predicted probability that plan \(p_i\) is indeed at its correct position \(y_i\).
It uses a softmax normalization across all possible positions \(j\) for that specific plan \(p_i\). 
The loss is the sum of the negative log probabilities over all plans in the list. 
Minimizing \(\mathcal{L}\) drives the model to increase the scores \(s_{i, y_i}\) corresponding to correct plan-position assignments relative to scores for incorrect positions (\(s_{i, j}\) where \(j \neq y_i\)).

This loss function is inherently position-aware due to its formulation.
The ground truth \(y_i\) directly incorporates the target rank position for each plan \(p_i\).
Furthermore, the softmax normalization is performed over the scores \(s_{ij}\) for all possible positions \(j\) for a fixed plan \(i\). 
This encourages the model to learn scores that distinguish the suitability of different rank positions for each plan, rather than focusing solely on the overall quality of the plan.
By explicitly using the target rank position \(y_i\) and normalizing across all potential positions, the loss compels the model to become sensitive to the desired location of each plan within the ranked list.

\subsection{Hybrid Test-time Plan Selection}
\label{sec:hybrid-decision-making}
To ensure robustness and mitigate risks associated with potentially inaccurate predictions from the learned model, especially for unfamiliar or Out-Of-Distribution (OOD) queries, CARPO incorporates a hybrid decision-making mechanism at query execution time. 
This approach leverages a trained classifier to assess the reliability of the learned model's output before executing a plan.

\noindent\underline{\textbf{OOD Detector Design.}}
The core of the hybrid approach is a trained OOD data detector, essentially a binary classifier.
It is designed to distinguish between input plan features that are similar to training data (\(D_{train}\)) and those that are substantially different.
Mathematically, let \(f: X \rightarrow [0, 1]\) represent this binary classifier, where \(X\) is the input feature space corresponding to top-ranked plan(s) of a certain query. 
The classifier assigns a probability score indicating the likelihood of the input belonging to the "in-distribution" class (by label 1 in our case). 
From this classifier, we can derive a confidence function \(g: X \rightarrow [0, 1]\), often defined as \((x) = \max(f(x), 1 - f(x))\).

The desired behavior of this classifier is that it exhibits a significant confidence gap between known and unknown inputs, or known as "training data memorization".
Mathematically, there exists thresholds \(\tau_{in}\) and \(\tau_{out}\) such that:
\begin{equation}
    g(x) \begin{cases}
\ge \tau_{in} & \text{if } x \in D_{train} \\
\le \tau_{out} & \text{if } x \notin D_{train}
\end{cases},
\end{equation}
where \(\tau_{in}\) is close to 1 and \(\tau_{out}\) is significantly smaller than \(\tau_{in}\). 
This confidence gap allows \((x)\) to effectively function as an indicator of whether the learned ranking model's prediction for input \(x\) is likely to be reliable.

\noindent\underline{\textbf{Hybrid Decision Workflow.}}
Recalling the observation that the performance difference among the top few candidate plans (e.g., k=3) for a query is often negligible, CARPO's hybrid decision protocol avoids rigidly selecting only the top-1 plan from the learned ranker. 
Instead, it employs a cascading evaluation strategy over the top-k ranked plans, integrating the OOD detector's confidence assessment.
Let \(\pi_M=(P_M^{(1)}, P_M^{(2)}, \cdots, P_M^{(k)}, \cdots)\) be the ranked list of candidate plans produced by CARPO's learned listwise ranker for a given query. 
Let \(P_{CBO}\) be the plan generated by the native Cost-Based Optimizer. 
The decision process to select the final plan for execution \(P_{final}\) can be summarized in Algorithm ~\ref{alg:cap}.
\begin{algorithm}
    \caption{Hybrid Decision Workflow}\label{alg:cap}
    \begin{algorithmic}[1] 
    \State Initialize $P_{final} \leftarrow NULL$
    \State Initialize $i \leftarrow 1$
    \While{$i \le k$ \textbf{and} $P_{final} = NULL$}
        \State Extract features $x^{(i)}$ corresponding to plan $P_M^{(i)}$
        \State Compute confidence $g(x^{(i)})$ using OOD detector
        \If{$g(x^{(i)}) \ge \tau_{in}$}
            \State $P_{final} \leftarrow P_M^{(i)}$ \Comment{In-distribution plan found}
        \Else
            \State $i \leftarrow i + 1$ \Comment{Check next plan}
        \EndIf
    \EndWhile
    \If{$P_{final} = NULL$} \Comment{No in-distribution plan in top-k}
        \State $P_{final} \leftarrow P_{CBO}$ \Comment{Fallback to CBO}
    \EndIf
    \State \Return{$P_{final}$}
\end{algorithmic}
\end{algorithm}

This protocol iteratively evaluates the confidence for the learned model's plans, starting from the highest-ranked one \(P_M^{(1)}\). 
It selects \(P_M^{(i)}\) (where \(i \leq k\)) if it is detected as in-distribution. 
The system only executes the fallback strategy if all top-k candidate plans suggested by the learned model are flagged as potentially unreliable (e.g., OOD or low confidence). 
In this case, it resorts to the reliable plan \(P_{CBO}\) provided by the native optimizer.
This approach, combining top-k consideration and OOD detection, enhances robustness. 
It allows CARPO to use high-quality learned plans when confident, while ensuring safe execution by using the CBO otherwise.

\section{Experiment}
In this section, we present a comprehensive experimental evaluation to validate the effectiveness of CARPO. 
Specifically, we seek to answer the following research questions:

\noindent
\textbf{RQ1:} \textbf{How effective} is CARPO's listwise ranking model compared to the native query optimizer of DBMS and SOTA learned optimizers? ($\vartriangleright$ Section~\ref{subsec:end_to_end_results})

\noindent
\textbf{RQ2:} \textbf{How adaptable} is the CARPO framework to different underlying plan embedding techniques (e.g., TreeCNN, TreeLSTM)? ($\vartriangleright$ Section~\ref{subsec:different_model})

\noindent
\textbf{RQ3:} \textbf{How indicative} is the performance distribution of top-ranked query plans for our hybrid top-$k$ plan selection strategy? ($\vartriangleright$ Section~\ref{subsec:case_study})



\subsection{Experiment Setup}
\noindent\textbf{\underline{Implementation and Environment.}}
All experiments were conducted on a Linux machine running Ubuntu 20.04 with kernel version 5.15.0-94-generic. 
The system is equipped with dual-socket Intel Xeon Platinum 8255C CPUs (96 threads total, 2.5GHz base frequency) and an NVIDIA GeForce RTX 2080 Ti GPU. 
PostgreSQL version 13.1 was used as the database engine.

\noindent\textbf{\underline{Compared Baselines.}}
We compare the performance of CARPO against two primary baselines: the native cost-based optimizer (CBO) of PostgreSQL and Lero, a representative pairwise learning-to-rank optimizer, to assess CARPO's effectiveness relative to alternative LTR approaches.

\noindent\textbf{\underline{Benchmarks.}}
We evaluate CARPO using standard query optimization benchmarks chosen to represent diverse workload characteristics. 
We utilize \ding{182}\textbf{TPC-H}\cite{tpch}, a widely-adopted benchmark representing decision support systems with complex queries involving multiple joins, aggregations, and filters. 
Additionally, we use \ding{183} \textbf{a selection of queries from STATS} benchmark\cite{skopal2011benchmarking}. 
For these queries, candidate execution plans are generated utilizing Lero's plan exploration strategy. 

To establish ground truth for training and evaluation, these generated candidate plans are executed beforehand locally to obtain their actual execution times. 
These times subsequently determine the correct performance ranking, which forms the basis for our listwise model training. 
The collected dataset, comprising queries along with their ranked plans and execution times, is then partitioned into training and testing sets using an 8:2 split.

\noindent\textbf{\underline{Evaluation Metrics.}}
We report two primary metrics to evaluate end-to-end performance:
\ding{182} \textbf{The cumulative execution time} on the test workload, calculated by summing the actual measured execution time of the plan selected by each optimizer for every query in the test set. 
\ding{183} \textbf{Top-$k$ accuracy}, which measures the percentage of test queries for which the plan selected by the optimizer is among the $k$ best-performing plans based on their actual measured execution times. 

\subsection{End-to-End Performance (RQ1)}\label{subsec:end_to_end_results}

\begin{figure}
    \centering
    \includegraphics[width=1\linewidth]{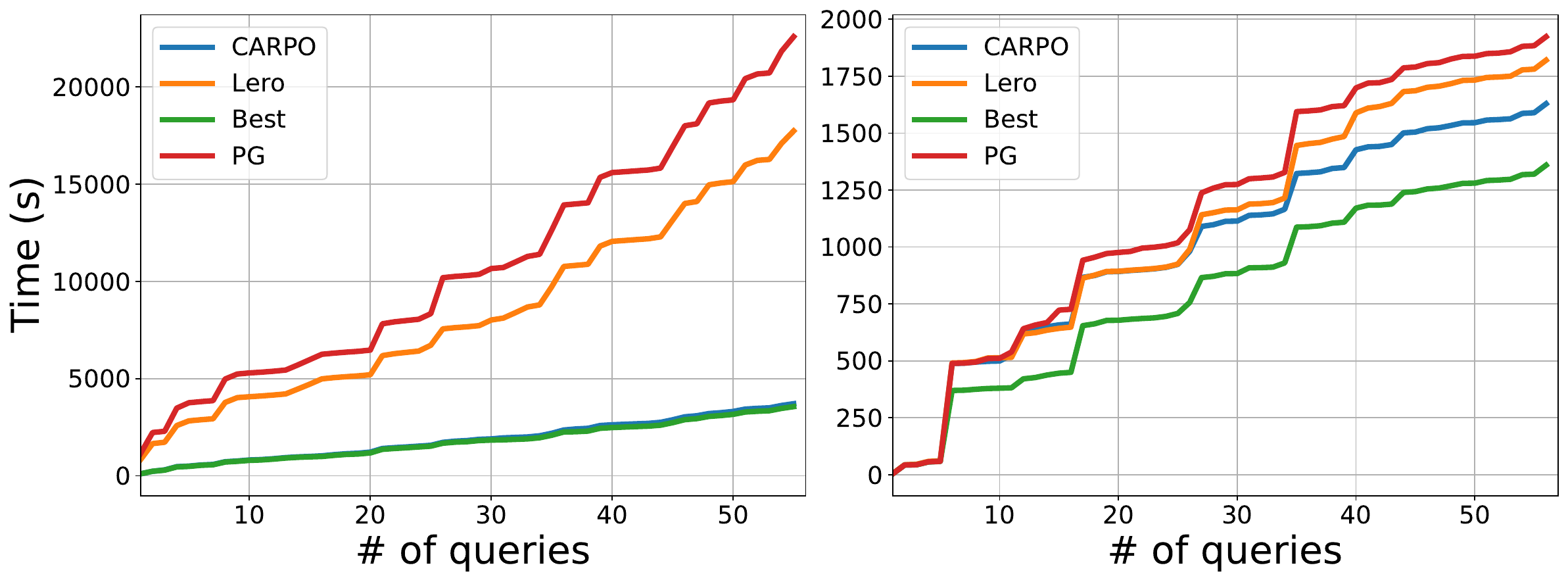}
    \caption{Cumulative execution time on the TPC-H test set (left) and STATS test set (right), comparing CARPO against Lero, PostgreSQL (Pg), and the optimal performance (Best).}
    \label{fig:overall-results}
\end{figure}

\noindent\textbf{\underline{Overall evaluation.}} 
To assess overall effectiveness, we first examine the cumulative execution time on the TPC-H and STATS test workloads, as depicted in Figure~\ref{fig:overall-results}. 
On TPC-H, CARPO achieves a total execution time of \textbf{3719.2 seconds}, significantly outperforming both the native PostgreSQL optimizer (pg) at 22577.9 seconds and the pairwise learned optimizer Lero at 17732.5 seconds.
It closely approaches the optimal performance represented by the best-found plans (bset) at 3559.9 seconds. 
This trend continues on the STATS benchmark, where CARPO records a cumulative time of \textbf{1628.7 seconds}, again notably lower than pg's 1923.1 seconds and Lero's 1819.9 seconds, and demonstrating competitiveness with the bset time of 1359.0 seconds. 
These results strongly indicate that CARPO's listwise ranking model is highly effective in selecting efficient query plans, leading to substantial overall workload time reductions compared to both traditional and pairwise learning-to-rank optimizers across both benchmarks.

\noindent\textbf{\underline{Model evaluation.}} 
Beyond cumulative time, we assess CARPO's effectiveness using top-$k$ accuracy.
Specifically, it reports the percentage of queries where the true best-performing plan is found within the top 1, 2, or 3 plans selected by our model. 
Results are shown in Table~\ref{tab:topk-results}.
On the TPC-H test set, CARPO achieves a top-1 accuracy of 74.54\% and a top-3 accuracy of 87.23\%, decisively outperforming Lero, which scores 3.63\% and 18.18\% for top-1 and top-3 accuracy, respectively.
For the STATS test set, CARPO registers a top-1 accuracy of 46.43\% and a top-3 accuracy of 80.36\%.
Lero's corresponding figures are 32.14\% and 72.5\%. 
These results demonstrate CARPO's significantly enhanced ability to identify optimal plans. 
The consistently high top-3 accuracy across both benchmarks indicates that CARPO’s top three predicted plans are highly likely to include the actual best execution strategy. 
This finding validates the effectiveness of incorporating a top-k consideration in our hybrid approach.

\begin{table}[ht]
\caption{Evaluation results of different query optimizers (CARPO, Lero, and native PostgreSQL) on TPC-H and STATS benchmarks.}
\label{tab:topk-results}
\begin{tabular}{c|c|ccc|c}
\hline
\textbf{\begin{tabular}[c]{@{}c@{}}Test\\ Dataset\end{tabular}}                           & \textbf{Model} & \textbf{\begin{tabular}[c]{@{}c@{}}Top 1\\ Rate\end{tabular}} & \textbf{\begin{tabular}[c]{@{}c@{}}Top 2\\ Rate\end{tabular}} & \textbf{\begin{tabular}[c]{@{}c@{}}Top 3\\ Rate\end{tabular}} & \textbf{\begin{tabular}[c]{@{}c@{}}Total\\ Time(ms)\end{tabular}} \\ \hline
\multirow{3}{*}{\begin{tabular}[c]{@{}c@{}}TPCH\\  Test Set\\  (8:2 Split)\end{tabular}}  & \textbf{CARPO} & \textbf{74.54\%}                                              & \textbf{83.63\%}                                              & \textbf{87.23\%}                                              & \textbf{3719.16}                                                  \\
                                                                                          & Lero           & 3.63\%                                                        & 3.63\%                                                        & 18.18\%                                                       & 17732.50                                                          \\
                                                                                          & PG             & \textbackslash{}                                              & \textbackslash{}                                              & \textbackslash{}                                              & 22577.87                                                          \\ \hline
\multirow{3}{*}{\begin{tabular}[c]{@{}c@{}}STATS\\  Test Set\\  (8:2 Split)\end{tabular}} & \textbf{CARPO} & \textbf{46.43\%}                                              & \textbf{69.64\%}                                              & \textbf{80.36\%}                                              & \textbf{1628.68}                                                  \\
                                                                                          & Lero           & 32.14\%                                                       & 62.5\%                                                        & 72.5\%                                                        & 1819.99                                                           \\
                                                                                          & PG             & \textbackslash{}                                              & \textbackslash{}                                              & \textbackslash{}                                              & 1923.09                                                           \\ \hline
\end{tabular}
\vspace{-5pt}
\end{table}

\noindent\textbf{\underline{Ablation Study.}}
To understand the impact of key hyperparameters on CARPO's performance and to validate our default settings, we conduct an ablation study on the \textbf{TPC-H benchmark}. 
We vary one hyperparameter at a time, keeping others at their empirically determined optimal values, and report top-$k$ accuracy and cumulative execution time.
Results are shown in Table~\ref{tab:ablation}.

\begin{itemize}[leftmargin=*]
\item 
\textbf{Effect of Model Latent Space Dimension $d_{model}$:}
We tested $d_{model}$ values of 32, 64, 128, and 256. CARPO achieved the best top-1 accuracy (72.72\%) at both 32 and 256, but performance dropped at 64 (45.45\%) and 128 (43.64\%). 
Top-3 accuracy showed a similar trend. 
This suggests 32 captures key features well without overfitting, while 256 benefits from higher capacity. 
Since execution time stayed stable (\~3718–3719ms), we choose $d_{model} = 32$ for its strong accuracy and compactness.

\item
\textbf{Effect of Transformer Layers: }
We varied the number of Transformer layers from 1 to 4. 
The optimal performance was clearly achieved with just 1 layer (74.54\% top-1, 87.27\% top-3 accuracy). 
Accuracy consistently decreased with additional layers (e.g., 4 layers yielded 41.82\% top-1). 
This suggests that for the TPC-H plan structures and features, extensive hierarchical processing might not be beneficial and could potentially lead to over-fitting. 
Cumulative time was stable for 1-3 layers (around 3719 ms), further supporting the choice of \textbf{a single, efficient layer}.

\item
\textbf{Effect of Learning Rate (lr):} 
Evaluating learning rates of 1E-03, 1E-04, 5E-05, and 1E-05, a rate of 5E-05 provided the best top-3 accuracy (87.27\%) and strong top-1 accuracy (63.63\%), with a marginally faster cumulative time (3712.96 ms). 
While 1E-03 yielded a higher top-1 accuracy (72.72\%), its top-3 accuracy was lower (76.36\%). 
The smaller learning rate of \textbf{5E-05} may allow for more stable convergence to a region of the parameter space that generalizes better across the top few ranks.
\end{itemize}

\subsection{Evaluation of CARPO+X (RQ2)}\label{subsec:different_model}
To address RQ2 concerning CARPO's adaptability to various plan embedding techniques, we compare two embedders TreeCNN and TreeLSTM.
Our finding demonstrates that while CARPO operates effectively with both, TreeLSTM consistently yields better performance. 
On TPC-H, TreeLSTM improves top-1 accuracy from 45.45\% to 72.72\% and top-3 from 58.18\% to 76.36\%; 
on STATS, it boosts top-1 from 46.43\% to 78.57\%, and reduces execution time from 1628.68ms to 1377.60ms. 
These results demonstrate that CARPO’s listwise architecture flexibly supports different embeddings while benefiting notably from richer representations like TreeLSTM.

\begin{table}[ht]
\caption{Evaluation results of different embedder(Tree-CNN, Tree-LSTM) on TPC-H and STATS benchmarks under the same set of training parameters.}
\label{tab:embedder-results}
\resizebox{\linewidth}{!}{
\begin{tabular}{c|cc|cc}
\hline
\textbf{Test Dataset}  & \multicolumn{2}{c|}{\textbf{TPC-H}} & \multicolumn{2}{c}{\textbf{STATS}} \\ \hline
\textbf{Plan Embedder} & TreeCNN      & TreeLSTM             & TreeCNN     & TreeLSTM             \\ \hline
\textbf{Top 1}         & 45.45\%      & 72.72\%              & 46.43\%     & 78.57\%              \\ \hline
\textbf{Top 2}         & 54.54\%      & 76.36\%              & 69.64\%     & 83.93\%              \\ \hline
\textbf{Top 3}         & 58.18\%      & 76.36\%              & 80.36\%     & 83.93\%              \\ \hline
\textbf{Time}          & 3719.16      & \textbf{3718.19}     & 1628.68     & \textbf{1377.60}     \\ \hline
\end{tabular}}
\vspace{-7pt}
\end{table}

\begin{table*}
\caption{
Ablation study results of CARPO.
\fcolorbox{white}{highlightmax}{Red} highlights the best results among chosen parameters.
}
\resizebox{\linewidth}{!}{
\begin{tabular}{c|cccc|cccc|cccc}
\hline
\multirow{2}{*}{\textbf{Parameters}} & \multicolumn{4}{c|}{\textbf{Latent Dimension}}          & \multicolumn{4}{c|}{\textbf{Layers}}              & \multicolumn{4}{c}{\textbf{Learning Rate}}                        \\ \cline{2-13} 
                                     & \textbf{32} & \textbf{64} & \textbf{128} & \textbf{256} & \textbf{1} & \textbf{2} & \textbf{3} & \textbf{4} & \textbf{1E-03} & \textbf{1E-04} & \textbf{5E-05} & \textbf{1E-05} \\ \hline
\textbf{Top 1}                       & \fcolorbox{white}{highlightmax}{72.72\%}     & 45.45\%     & 43.64\%      & 72.72\%      & \fcolorbox{white}{highlightmax}{74.54\%}    & 45.45\%    & 45.45\%    & 41.82\%    & \fcolorbox{white}{highlightmax}{72.72\%}        & 45.45\%        & 63.63\%        & 41.82\%        \\ \hline
\textbf{Top 2}                       & \fcolorbox{white}{highlightmax}{76.36\%}     & 54.54\%     & 47.27\%      & 76.36\%      & \fcolorbox{white}{highlightmax}{83.63\%}    & 54.54\%    & 54.54\%    & 47.27\%    & 76.36\%        & 54.54\%        & \fcolorbox{white}{highlightmax}{83.63\%}        & 47.27\%        \\ \hline
\textbf{Top 3}                       & \fcolorbox{white}{highlightmax}{76.36\%}     & 58.18\%     & 47.27\%      & 76.36\%      & \fcolorbox{white}{highlightmax}{87.27\%}    & 58.18\%    & 58.18\%    & 47.27\%    & 76.36\%        & 58.18\%        & \fcolorbox{white}{highlightmax}{87.27\%}        & 47.27\%        \\ \hline
\textbf{Time}                        & \fcolorbox{white}{highlightmax}{3718.19}     & 3719.16     & 3718.19      & 3718.19      & \fcolorbox{white}{highlightmax}{3719.16}    & 3719.16    & 3719.16    & 3730.69    & 3718.19        & 3719.16        & \fcolorbox{white}{highlightmax}{3712.96}        & 3712.96        \\ \hline
\end{tabular}
}
\label{tab:ablation}
\end{table*}

\subsection{In-depth Case Studies (RQ3)}\label{subsec:case_study}
This subsection addresses how the performance distribution of top-ranked query plans informs our hybrid top-$k$ selection strategy (RQ3). 
We investigate the common observation that multiple leading candidate plans for a query frequently exhibit negligible differences in actual execution time. 
Through specific case studies, we aim to visually and analytically demonstrate this characteristic.
These examples will illustrate how the proximity in performance of top-$k$ plans validates the design rationale behind CARPO's robust hybrid decision protocol, which considers several high-quality candidates before resorting to a fallback.

\begin{figure}[!ht]
    \centering
    \includegraphics[width=0.8\linewidth]{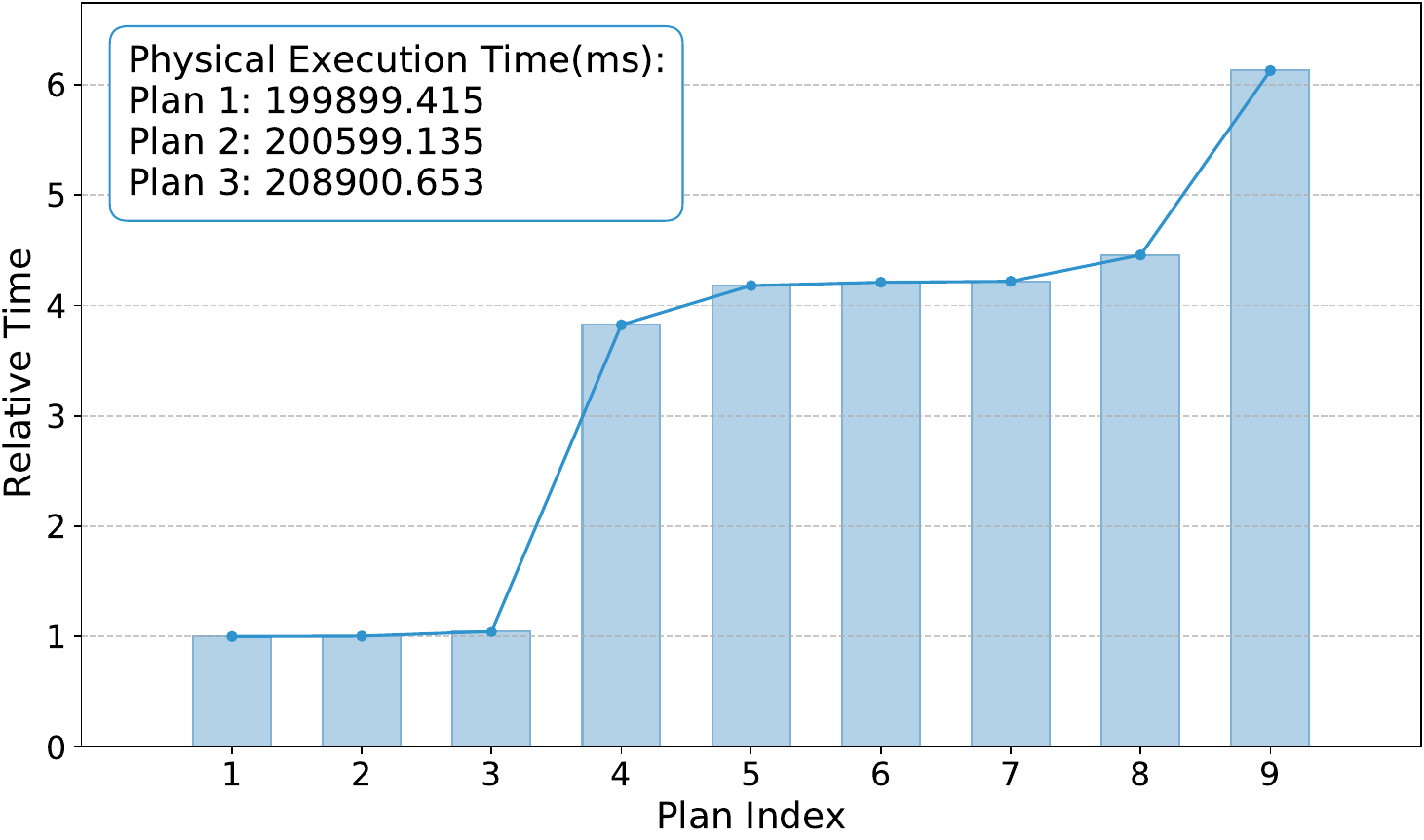}
    \caption{Example Query 1 from TPC-H benchmark.}
    \label{fig:case 1}
\end{figure}

\noindent\textbf{\underline{Case 1: Typical Case in TPC-H Benchmark.}}
Figure~\ref{fig:case 1} illustrates a common performance distribution observed for candidate plans of TPC-H Query.
The execution times of the top-ranked plans (e.g., Plans 1, 2, and 3 in the example, with times of 199.9s, 200.6s, and 208.9s respectively) are notably clustered, exhibiting minimal relative differences. 
This tight grouping at the top contrasts with later-ranked plans, which generally operate at a significantly higher latency baseline. 
This characteristic underscores the rationale for CARPO's hybrid top-$k$ selection: if the highest-ranked plan is flagged as OOD, CARPO can confidently select another plan from this high-performing cluster (like Plan 2 or 3) without substantial performance degradation, thereby ensuring robustness while leveraging the learned model's ability to identify this competitive set.

\begin{figure}[ht]
    \centering
    \includegraphics[width=0.8\linewidth]{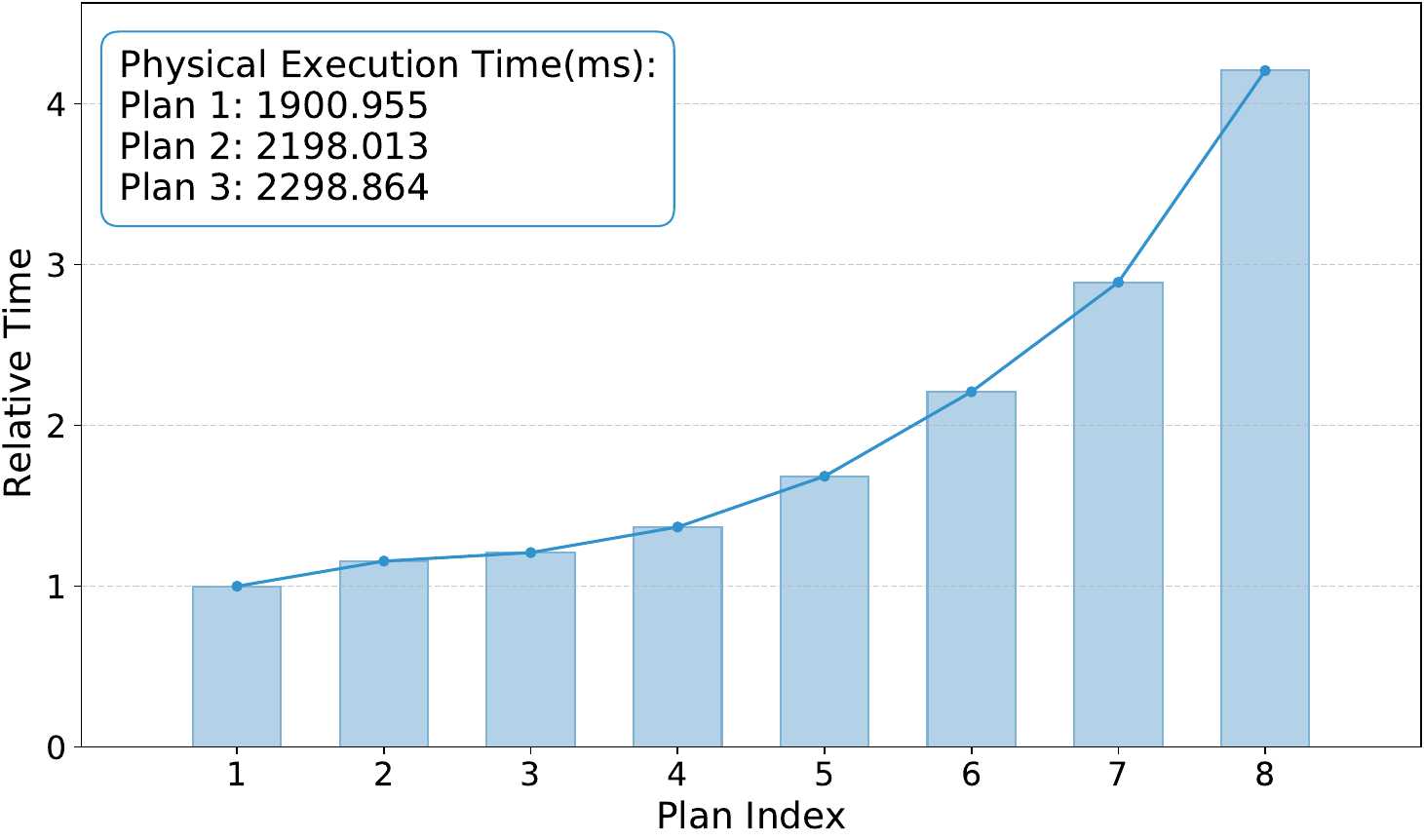}
    \caption{Example Query 2 from STATS benchmark.}
    \label{fig:case 2}
    \vspace{-10pt}
\end{figure}

\noindent\textbf{\underline{Case 2: Typical Case in STATS Benchmark.}}
Figure~\ref{fig:case 2} presents the execution time distribution for a representative query from the STATS benchmark. 
Unlike the more clearly partitioned performance tiers often seen in TPC-H (as illustrated in Case 1), the plan execution times for this STATS query demonstrate a more gradual increase.
However, the crucial top-$k$ effect is still evident: the initial set of plans (Plans 1-3 in Figure~\ref{fig:case 2}, with times of 1.9s, 2.2s, 2.3s, respectively) show relatively small increments in latency. 
While subsequent plans exhibit progressively higher execution times, the initial cluster of good-performing plans remains distinct. 
This observation from STATS further reinforces the value of CARPO's hybrid top-$k$ strategy. 
Even when a sharp performance drop-off isn't present, the ability to select among several closely performing leading candidates provides a robust path to efficient query execution.

\subsection{Discussion on Hybrid Decision Block}
\label{sec:discuss-hybird}

\noindent\textbf{\underline{Tie Prediction Scenario.}}
A noteworthy scenario during CARPO's prediction phase is the occurrence of ties, where multiple candidate plans receive the highest or indistinguishable scores.
These ties often correspond to plans with very similar actual execution times, suggesting that the embedding and ranking model effectively groups structurally or semantically similar plans—despite not explicitly predicting latency.
While this correlation is encouraging, it introduces a risk: if a suboptimal plan is mistakenly included among top-ranked ties, random selection could lead to poor performance.
This is where the hybrid decision block (Section \ref{sec:hybrid-decision-making}) plays a crucial role.
When CARPO encounters a tie, each top-ranked plan (up to $k$ candidates) is passed through the Out-Of-Distribution (OOD) detector.
If the selected plan scores low in confidence, the hybrid mechanism falls back to the PostgreSQL CBO’s choice.
This safeguard ensures that even tied selections are vetted, reducing the risk of executing unreliable plans.

\begin{figure}[ht]
\centering
\includegraphics[width=1\linewidth]{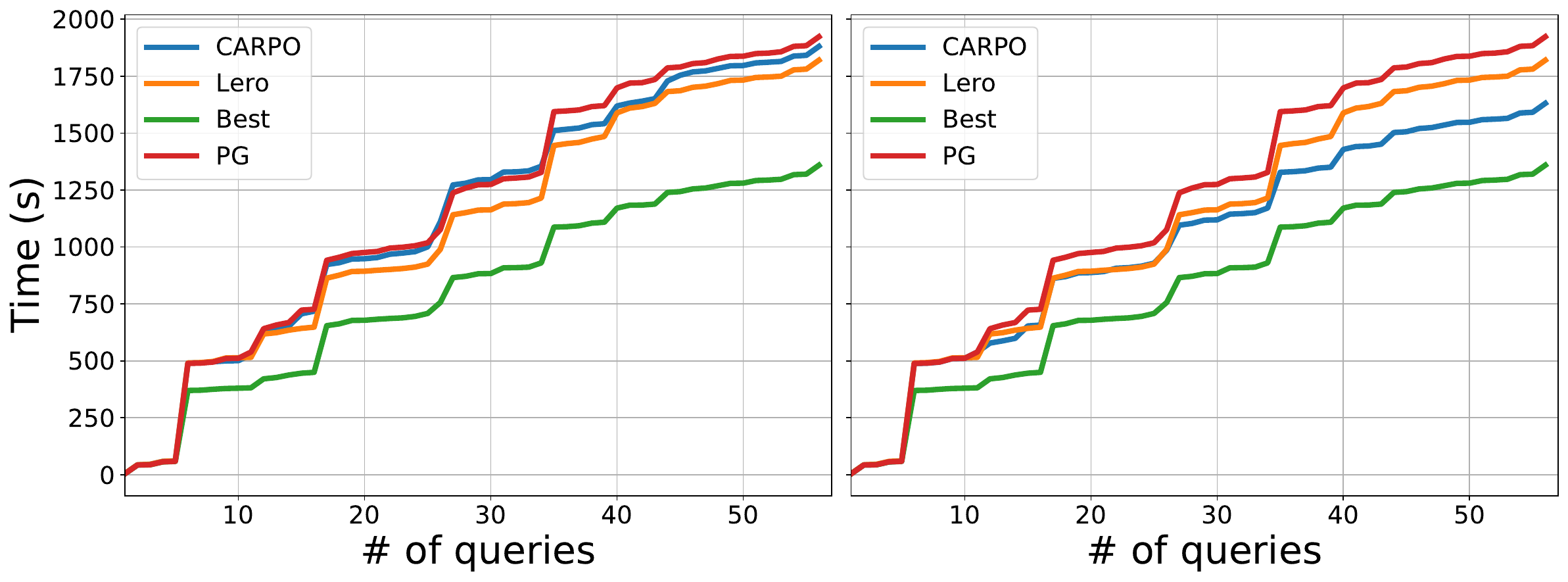}
\caption{
Effect of the hybrid fallback strategy on CARPO's performance on the STATS test set. 
\textbf{Left:} Performance without fallback.
\textbf{Right:} Performance with fallback.
}
\label{fig:hybrid_fallback_effect}
\end{figure}

\noindent\textbf{\underline{Impact of Hybrid Fallback.}}
The efficacy of this hybrid decision block in practical scenarios is further illustrated in Figure~\ref{fig:hybrid_fallback_effect}. 
This comparison on the STATS test set demonstrates that enabling the fallback strategy significantly improves CARPO's robustness. 
By intervening when the learned model's top predictions are unreliable, the hybrid mechanism constrains potential performance regressions and leads to more consistent and predictable query execution times, effectively preventing severe slowdowns that might occur if relying solely on potentially erroneous model outputs.

\section{Conclusion}
In this paper, we introduced CARPO, a novel listwise learning-to-rank framework designed to address the inherent limitations of traditional cost-based optimizers and existing pairwise learned query optimizers. 
CARPO uniquely employs a Transformer-based architecture to holistically evaluate candidate query plans, capturing inter-plan dependencies, and integrates a robust hybrid decision mechanism with Out-Of-Distribution detection and a top-$k$ fallback strategy to ensure both performance and reliability. 
Our comprehensive experiments on TPC-H and STATS benchmarks demonstrate that CARPO significantly outperforms the native PostgreSQL CBO and the pairwise optimizer Lero in both cumulative execution time and top-$k$ plan accuracy. 
Furthermore, we validated CARPO's adaptability to different plan embedders like TreeCNN and TreeLSTM. 
Through case studies and performance analysis, we confirmed that its hybrid top-$k$ strategy is well-justified by the typical performance distribution of query plans and the efficacy of the fallback mechanism.
CARPO thus establishes a more effective and robust approach to learned query optimization.

%
\bibliographystyle{ACM-Reference-Format}
\bibliography{main}






\end{document}